\documentclass[prd,preprint,aps,showpacs]{revtex4}
\usepackage{graphics}
\newcommand{\newc}{\newcommand}
\newc{\gsim}{\lower.7ex\hbox{$\;\stackrel{\textstyle>}{\sim}\;$}}
\newc{\lsim}{\lower.7ex\hbox{$\;\stackrel{\textstyle<}{\sim}\;$}}
\newc{\gev}{\,{\rm GeV}}
\newc{\mev}{\,{\rm MeV}}
\newc{\ev}{\,{\rm eV}}
\newc{\kev}{\,{\rm keV}}
\newc{\tev}{\,{\rm TeV}}
\newc{\mz}{m_Z}
\newc{\mpl}{M_{Pl}}
\newc{\chifc}{\chi_{{}_{\!F\!C}}}
\newc\order{{\cal O}}
\newc\CO{\order}
\newc\CL{{\cal L}}
\newc\CY{{\cal Y}}
\newc\CH{{\cal H}}
\newc\CM{{\cal M}}
\newc\CF{{\cal F}}
\newc\CD{{\cal D}}
\newc\CN{{\cal N}}
\newc{\eps}{\epsilon}
\newc{\re}{\mbox{Re}\,}
\newc{\im}{\mbox{Im}\,}
\newc{\invpb}{\,\mbox{pb}^{-1}}
\newc{\invfb}{\,\mbox{fb}^{-1}}
\newc{\yddiag}{{\bf D}}
\newc{\yddiagd}{{\bf D^\dagger}}
\newc{\yudiag}{{\bf U}}
\newc{\yudiagd}{{\bf U^\dagger}}
\newc{\yd}{{\bf Y_D}}
\newc{\ydd}{{\bf Y_D^\dagger}}
\newc{\yu}{{\bf Y_U}}
\newc{\yud}{{\bf Y_U^\dagger}}
\newc{\ckm}{{\bf V}}
\newc{\ckmd}{{\bf V^\dagger}}
\newc{\ckmz}{{\bf V^0}}
\newc{\ckmzd}{{\bf V^{0\dagger}}}
\newc{\X}{{\bf X}}
\newc{\bbbar}{B^0-\bar B^0}

\newc{\sgn}{\mbox{sgn}\,}
\newc{\m}{{\bf m}}
\newc{\msusy}{M_{\rm SUSY}}
\newc{\munif}{M_{\rm unif}}
\newc{\slepton}{{\tilde\ell}}
\newc{\Slepton}{{\tilde L}}
\newc{\sneutrino}{{\tilde\nu}}
\newc{\selectron}{{\tilde e}}
\newc{\stau}{{\tilde\tau}}
\def\beq{\begin{equation}}
\def\eeq{\end{equation}}
\def\bea{\begin{eqnarray}}
\def\eea{\end{eqnarray}}
\newc{\ie}{{\it i.e.}}          \newc{\etal}{{\it et al.}}
\newc{\eg}{{\it e.g.}}          \newc{\etc}{{\it etc.}}
\newc{\cf}{{\it c.f.}}
\def\beqn{\begin{eqnarray}}
\def\eeqn{\end{eqnarray}}
\relax
\newcommand{\ba}[1]{\begin{array}{#1}}
\def\ea{\end{arroy}}

\def\beq{\begin{equation}}
\def\eeq{\end{equation}}
\def\bea{\begin{array}}
\def\eea{\end{array}}

\def\to{\rightarrow}

\def\[{\left[}
\def\]{\right]}
\def\({\left(}
\def\){\right)}

\def\U1em{{U(1)_{\rm em}}}
\def\to{\rightarrow}

\def\CL{{\cal C}_L}

\def\sq2{\sqrt{2}}

\def\End{\end{document}}

\def\Dsl{\,\raise.65ex\hbox{/}\mkern-03.5mu D} 
\def\delsl{\raise.15ex\hbox{/}\kern-.57em\partial}
\def\Ksl{\hbox{/\kern-.6700em\rm K}}
\def\Asl{\hbox{/\kern-.1500em \rm A}}
\def\Qsl{\hbox{/\kern-.6000em\rm Q}}
\def\gradsl{\hbox{/\kern-.6501em$\nabla$}}
\def\bar#1{\overline{#1}}

\begin{document}
\draft
\title{Charged Higgs boson contribution to $\nu_{\tau} {\cal N} \to
\tau^- X$ for very large $\tan\beta$ in the two Higgs doublet model
with UHE-neutrinos}
\author{A. Rosado}
\address{Instituto de F\'{\i}sica, BUAP. Apdo. Postal J-48, C.P.
12570 Puebla, Pue., M\'exico}
\date{\today}
\begin{abstract}
We study the deep inelastic process $\nu_{\tau} + {\cal N} \to
\tau^{-} + X$ (with ${\cal N} \equiv (n+p)/2$ an isoscalar
nucleon), in the context of the two Higgs doublet model {\it type
two} (2HDM(II)). In particular, we discuss the contribution to the
total cross section of diagrams, in which a charged Higgs boson is
exchanged. We show that for large values of $\tan\beta$ such
contribution for an inclusive dispersion generated through the
collision of an ultrahigh energy tau-neutrino on a target nucleon
can reach up to $57\%$ of the value of the contribution of the
$W^+$ exchange diagrams (i.e. can reach up to $57\%$ of the
standard model (SM) prediction) and could permit to distinguish
between the SM and the 2HDM(II) predictions at the Pierre Auger
Observatory.
\end{abstract}
\maketitle

\setcounter{footnote}{0}
\setcounter{page}{2}
\setcounter{section}{0}
\setcounter{subsection}{0}
\setcounter{subsubsection}{0}

\narrowtext

\section{Introduction.}

Although the Standard Model (SM) \cite{stanmod}, of the strong and
electroweak interactions describes correctly Particle Physics at
present energies, one of its basic ingredients, the scalar Higgs
sector, still remains untested. In the SM, the Higgs sector
consists of a single $SU(2)$ doublet, and after spontaneous
symmetry breaking (SSB) it remains a physical state, the Higgs
boson ($h^0_{sm}$), whose mass is not predicted in the theory. On
the other hand, the SM is not expected to be the ultimate
theoretical structure responsible for electroweak symmetry
breaking (EWSB) \cite{HHG,TSM}. One of the most simple extensions
of the SM is the so called two Higgs doublet model (2HDM). There
are four classes of 2HDM which naturally avoid tree-level
flavor-changing neutral currents that can be induced by Higgs
boson exchange \cite{Barger:1989fj}. These models include a Higgs
sector with two scalar doublets, which give masses to the up and
down-type fermions as well as the gauge bosons. Particularly
interesting is the model II, in this model one of the Higgs scalar
doublet couples to the up-components of isodoublets while the
second one to the down-components. Model II is that one which is
present in SUSY theories. The 2HDM(II) has a rich structure and
predicts interesting phenomenology \cite{HHG}. The physical
spectrum consists of two neutral CP-even states ($h^{0},H^{0}$)
and one CP-odd ($A^{0} $), as well as a pair of charged scalar
particles ($H^{\pm }$). The charged Higgs boson in this model has
the following lower mass limits \cite{Heister:2002ev} and
\cite{Krawczyk:2004na}:
\begin{equation}\label{bound1}
M_{H^{\pm}} > 79.3 \, \mbox{GeV}\\
\end{equation}
and
\begin{equation}\label{bound2}
M_{H^{\pm}} \gsim 1.71 \, \tan\beta \, \mbox{GeV}\\
\end{equation}

Large-scale neutrino telescopes \cite{telescopes} have as a main
goal the detection of ultrahigh-energy (UHE) cosmic neutrinos
($E_\nu \geq 10^{12}$ eV) produced outside the atmosphere
(neutrinos produced by galactic cosmic rays interacting with
interstellar gas, and extragalactic neutrinos)
\cite{uhe-neut,Gandhi:1998ri}. UHE neutrinos can be detected by
observing long-range muons and tau-leptons decays produced in
charged-current neutrino-nucleon interactions. UHE tau-neutrinos
are generated through neutrino oscillations
\cite{Blanch1,Bertou:2001vm}. The detection of UHE neutrinos will
provide us with the possibility to observe $\nu {\cal
N}$-collisions with a neutrino energy in the range $10^{12}$ eV
$\leq E_\nu \leq 10^{21}$ eV and a target nucleon at rest. An
enlightening discussion on UHE neutrino interactions is given by
R. Gandhi {\it et al.} \cite{Gandhi:1998ri}.

We discuss in this paper the cross section of the deep inelastic
process $\nu_{\tau} + {\cal N} \to \tau^{-} + X$ (${\cal N} \equiv
(n+p)/2$ an isoscalar nucleon), in the context of the SM and the
2HDM(II) by using the parton model \cite{partmod} with the parton
distribution functions reported by J. Pumplin {\it et al.}
\cite{pumplin}. We use the CTEQ PDFs provided in a $n_f=5$ active
flavors scheme. Our aim is to calculate how large can be the
contribution of diagrams, in which a charged Higgs boson is
exchanged, to the total cross section of the mentioned inclusive
process in the frame of the 2HDM(II). In the 2HDM(II) the
couplings of the down-type quarks and charged leptons are
proportional to $m_{f} \times \tan\beta$. Hence, for large
$\tan\beta$ the contribution of $H^{\pm}$-exchange diagrams will
be maximal in this model.

\section{The cross section for the inclusive process $\nu_{\tau} + {\cal N} \to
\tau^- + X$}

\subsection{The differential cross section for the process
$\nu_{\tau} + {\cal N} \to \tau^- + X$ in the SM}

\begin{figure}[floatfix]
\begin{center}
\includegraphics{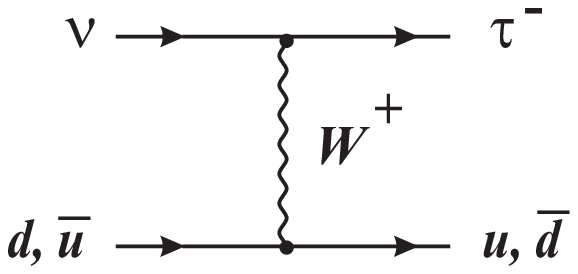}
\caption{Diagrams which contribute at the quark level to the
process $\nu_{\tau} + {\cal N} \to \tau^- + X$ at the lowest order
in $\alpha$ in the SM ($d$ stands for $d$-, $s$- and $b$-quark;
and $u$ stands for $u$- and $c$-quark).} \label{fig:figsm}
\end{center}
\end{figure}

The differential cross section for the inclusive reaction
\begin{equation}\label{procsm}
\nu_{\tau}(p) + {\cal N}(P_{\cal N}) \to  \tau^-(p') + X \, ,
\end{equation}
where ${\cal N} \equiv (n+p)/2$ is an isoscalar nucleon, at the
lowest order in $\alpha$ in the frame of the SM (see
Fig.~\ref{fig:figsm}) is given as follows \cite{dis-sm}:
\begin{equation}\label{dcssm}
\frac{d^2\sigma_{sm}}{dxdy}= \frac{2 G^2_F M E_{\nu}}{\pi} \left(
\frac {M^2_W}{Q^2+M^2_W} \right)^2 [x q_W(x,Q^2) + x
\bar{q}_W(x,Q^2) (1-y)^2] \, ,
\end{equation}
where $M$ stands for the nucleon mass, $Q^2$, $x$ and $y$ are
defined as usual
\begin{equation}\label{invar}
s=(p+P_{\cal N})^2 \, , \hspace{1cm} Q^2=-(p-p')^2 \, ,
\hspace{1cm} \nu=P_{\cal N}(p-p') \, , \\
\end{equation}
\noindent and
\begin{equation}\label{dimless}
x=\displaystyle\frac{Q^2}{2\nu} \, , \hspace{1cm}
y=\displaystyle\frac{2\nu}{s} \, . \hspace{1cm}
\end{equation}
The quantities $q_W(x,Q^2)$ and $\bar{q}_W(x,Q^2)$ are given as
\begin{eqnarray}\label{fqw}
q_W(x,Q^2) &=& \frac{u_v(x,Q^2) + d_v(x,Q^2)}{2}+\frac{u_s(x,Q^2)
+ d_s(x,Q^2)}{2} + s_s(x,Q^2) + b_s(x,Q^2)
\, , \nonumber\\
\bar{q}_W(x,Q^2) &=& \frac{u_s(x,Q^2) + d_s(x,Q^2)}{2} +
c_s(x,Q^2) \, ,
\end{eqnarray}
where the valence and sea parton distribution functions (PDFs),
$q_v(x,Q^2)$ and $q_s(x,Q^2)$, can be expressed as
\begin{eqnarray}\label{fqprot}
u_v(x,Q^2) &=& u(x,Q^2) - \bar{u}(x,Q^2) \, , \nonumber\\
d_v(x,Q^2) &=& d(x,Q^2) - \bar{d}(x,Q^2) \, , \nonumber\\
u_s(x,Q^2)&=&\bar{u}(x,Q^2) \, , \nonumber\\
d_s(x,Q^2)&=&\bar{d}(x,Q^2) \, , \nonumber\\
c_s(x,Q^2) &=& c(x,Q^2)=\bar{c}(x,Q^2) \, , \nonumber\\
s_s(x,Q^2) &=& s(x,Q^2)=\bar{s}(x,Q^2) \, , \nonumber\\
b_s(x,Q^2) &=& b(x,Q^2)=\bar{b}(x,Q^2) \, ,
\end{eqnarray}
where the PDFs $q(x,Q^2)$ describe the quark $q$ content of the
proton.

In the case of the standard model the couplings of the fermions to
the $W^{\pm}$ boson are given by the lagrangian
\begin{equation}\label{lagwsm} {\cal L}=-\frac{g}{\sqrt{2}}
\sum_{(f_u,f_d)}\left\{ \left( \bar{f}_u \gamma^{\mu}
\frac{1-\gamma_5}{2} f_d \right) W^+_{\mu} + \left( \bar{f}_d
\gamma^{\mu} \frac{1-\gamma_5}{2} f_u \right) W^-_{\mu} \right\}
\, ,
\end{equation}
where $f_u$ and $f_d$ stand for the up- and down-components of the
fermion doublet.

\subsection{The differential cross section for the process
$\nu_{\tau} + {\cal N} \to \tau^- + X$ in the 2HDM(II)}

\begin{figure}[floatfix]
\begin{center}
\includegraphics{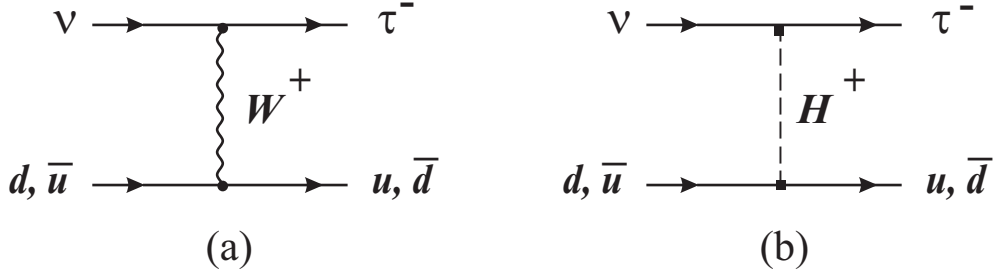}
\caption{Diagrams which contribute at the quark level to the
process $\nu_{\tau} + {\cal N} \to \tau^- + X$ at the lowest order
in $\alpha$ in the 2HDM ($d$ stands for $d$-, $s$- and $b$-quark;
and $u$ stands for $u$- and $c$-quark).} \label{fig:fig2hdm}
\end{center}
\end{figure}

The differential cross section for the inclusive reaction
(\ref{procsm}), at the lowest order in $\alpha$ in the frame of
the 2HDM(II) (see Fig.~\ref{fig:fig2hdm}), can be written as
follows:
\begin{equation}\label{dcs2hdm1}
\frac{d^2\sigma_{2hdm}}{dxdy}=\frac{d^2\sigma_{sm}}{dxdy} +
\frac{d^2\sigma_{H^+}}{dxdy} \, ,
\end{equation}
where for large $\tan\beta$
\begin{equation}\label{dcs2hdmh}
\frac{d^2\sigma_{H^+}}{dxdy}= \frac{G^2_F M E_{\nu}}{2 \pi} \frac
{m^2_{\tau} M^2_W \tan^4\beta}{(Q^2+M^2_{H^{\pm}})^2} y^2 [x
q_H(x,Q^2) + x \bar{q}_H(x,Q^2)] \, ,
\end{equation}
where $Q^2$, $x$ and $y$ are defined in (\ref{invar}) and
(\ref{dimless}) and $M$ stands for the nucleon mass. The
quantities $q_H(x,Q^2)$ and $\bar{q}_H(x,Q^2)$ are given as
follows
\begin{eqnarray}
q_H (x,Q^2) &=& \frac{m^2_d}{M^2_W} \left( \frac{u_v(x,Q^2) +
d_v(x,Q^2)}{2} +
\frac{u_s(x,Q^2) + d_s(x,Q^2)}{2} \right) \nonumber\\
&& + \frac{m^2_s}{M^2_W} \, s_s(x,Q^2) + \frac{m^2_b}{M^2_W} \,
b_s(x,Q^2) \nonumber\\
\bar{q}_H (x,Q^2) &=& \frac{m^2_d}{M^2_W} \left( \frac{u_s(x,Q^2)
+ d_s(x,Q^2)}{2} \right) + \frac{m^2_s}{M^2_W} \, c_s(x,Q^2) \, .
\end{eqnarray}

In the case of the 2HDM(II) the couplings of the fermions to the
$W^{\pm}$ boson are given, in a similar way as in the SM
\cite{HHG}, by the lagrangian in Eq.(\ref{lagwsm}). On the other
side, taking the elements of the CKM-matrix $V_{ij}= \delta_{ij}$,
the couplings of the fermions to the $H^{\pm}$ boson are given by
the lagrangian \cite{HHG}
\begin{equation}\label{lagh2hdm}
{\cal L}=\frac{g}{ M_W} \left\{ m_{\tau} \tan\beta \left(
\bar{\nu} \frac{1+\gamma_5}{2} \tau \right) + m_{u} \cot\beta
\left( \bar{u} \frac{1-\gamma_5}{2} d \right) + m_{d} \tan\beta
\left( \bar{u} \frac{1+\gamma_5}{2}d \right) \right\}H^+ + h.c.
\end{equation}

\section{Results for deep inelastic $\nu_{\tau} {\cal N}$ in the SM and
the 2HDM(II)}

We present results for the case of unpolarized deep inelastic
process $\nu_{\tau} + {\cal N} \to \tau^- + X$ with a neutrino
energy in the range $10^{14}$ eV  $\leq E_{\nu} \leq 10^{20}$ eV
and the nucleon at rest, $\it{i.e.}$ a target nucleon. We take
$10^{14}$ eV $\leq E_{\nu}$ to make possible to neglect all
fermion masses with respect to the total energy $s=2 m_{\cal N}
E_{\nu}$, even the top quark mass. We take cuts of $\sim 2$
GeV$^2$ and 10 GeV$^2$ for $Q^2$ and the invariant mass $W$,
respectively. We have checked numerically that the total cross
section rates do not depend on the choice of the cuts on the
momentum transfer square $Q^2$, when they take on values of a few
GeV$^2$. The reason of this fact is that the propagators involved
in the cross section calculation go as $1/(M^2_W+Q^2)$ and
$1/(M^2_{H^{\pm}}+Q^2)$

We perform our numerical calculations taking for the quark masses:
$m_u=4$ MeV, $m_d=8$ MeV, $m_c=1.5$ GeV, $m_s=150$ MeV, $m_b=4.9$
GeV and $m_t=174$ GeV \cite{partdata}. For the evaluation of the
$H^+ \tau^- \nu_{\tau}$ coupling we take $m_{\nu_{\tau}}=0$ and
$m_{\tau}=1,777$ MeV.

\begin{figure}[floatfix]
\begin{center}
\includegraphics{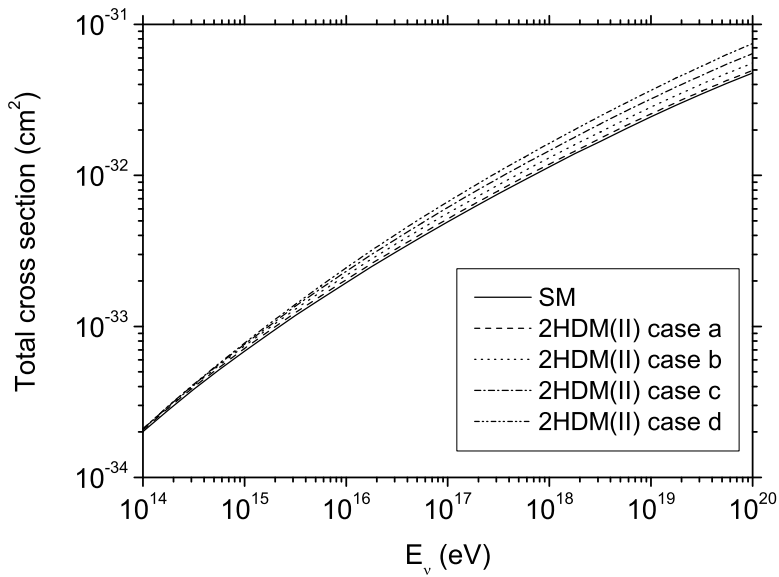}
\caption{Total cross section as a function of $E_{\nu}$ in the
range $10^{14}$ eV $\leq E_{\nu} \leq 10^{20}$ eV, with $E_{\cal
N}=m_{\cal N}$. We compare $\sigma^{tot}_{sm}$ with
$\sigma^{tot}_{2hdm}$ by taking: (a) $M_{H^{\pm}}=100$ GeV and
$\tan\beta=50$; (b) $M_{H^{\pm}}=200$ GeV and $\tan\beta=100$; (c)
$M_{H^{\pm}}=300$ GeV and $\tan\beta=150$; (d) $M_{H^{\pm}}=400$
GeV and $\tan\beta=200$. }\label{fig:sigmatot}
\end{center}
\end{figure}

We take $M_{W^+}=80.4$ GeV for the mass of the charged boson $W^+$
\cite{partdata} and show in Fig.~\ref{fig:sigmatot} our numerical
results for the total cross section as a function of $E_{\nu}$. We
compare the results for the $\sigma^{tot}_{sm}$ with those
obtained for the $\sigma^{tot}_{2hdm}$ by taking: (a)
$M_{H^{\pm}}=100$ GeV and $\tan\beta=50$; (b) $M_{H^{\pm}}=200$
GeV and $\tan\beta=100$; (c) $M_{H^{\pm}}=300$ GeV and
$\tan\beta=150$; (d) $M_{H^{\pm}}=400$ GeV and $\tan\beta=200$.
Since in all cases $M_{H^{\pm}}=2 \, \tan\beta$ GeV, then the
conditions (\ref{bound1}) and (\ref{bound2}) are fulfilled.
Further, based on the discussions on $\tan\beta$ given in Refs.
\cite{Roy:2005yu} and \cite{Baer:2002hf} we restrict ourselves to
take $\tan\beta \leq 200$.

\begin{figure}[floatfix]
\begin{center}
\includegraphics{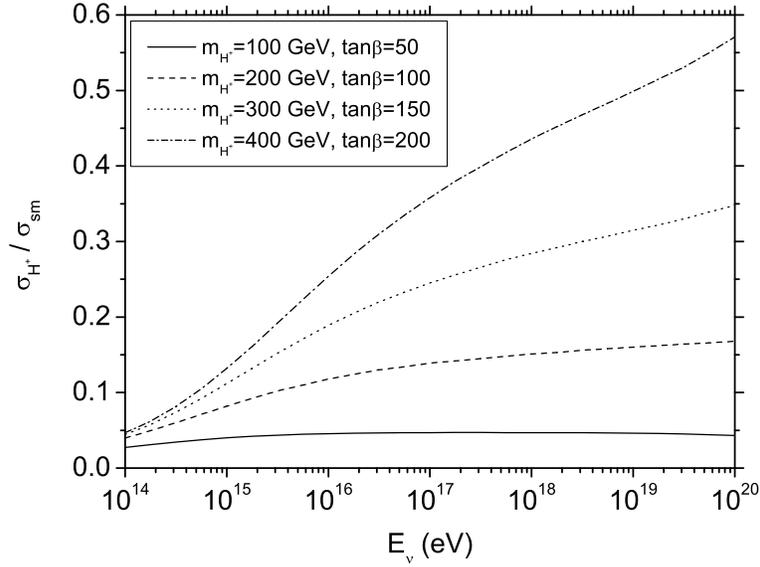}
\caption{$\sigma^{tot}_{H^+}(\nu_{\tau} + {\cal N} \to \tau^{-} +
X)/\sigma^{tot}_{sm}(\nu_{\tau} + {\cal N} \to \tau^{-} + X)$ as a
function of $E_{\nu}$ in the range $10^{14}$ eV $\leq E_{\nu} \leq
10^{20}$ eV, with $E_{\cal N}=m_{\cal N}$ for the cases: (a)
$M_{H^{\pm}}=100$ GeV and $\tan\beta=50$; (b) $M_{H^{\pm}}=200$
GeV and $\tan\beta=100$; (c) $M_{H^{\pm}}=300$ GeV and
$\tan\beta=150$; (d) $M_{H^{\pm}}=400$ GeV and
$\tan\beta=200$.}\label{fig:ratio}
\end{center}
\end{figure}

Further, we present in Fig.~\ref{fig:ratio} our results for the
ratio $\sigma^{tot}_{H^+}(\nu_{\tau} + {\cal N} \to \tau^{-} +
X)/\sigma^{tot}_{sm}(\nu_{\tau} + {\cal N} \to \tau^{-} + X)$ as a
function of $E_{\nu}$ for the cases: (a) $M_{H^{\pm}}=100$ GeV and
$\tan\beta=50$; (b) $M_{H^{\pm}}=200$ GeV and $\tan\beta=100$; (c)
$M_{H^{\pm}}=300$ GeV and $\tan\beta=150$; (d) $M_{H^{\pm}}=400$
GeV and $\tan\beta=200$. In particular, we have gotten
$\sigma^{tot}_{H^+}/\sigma^{tot}_{sm} = 0.57$ for $E_{\nu} =
10^{20}$ eV, $M_{H^{\pm}}=400$ GeV and $\tan\beta=200$.

\begin{figure}[floatfix]
\begin{center}
\includegraphics{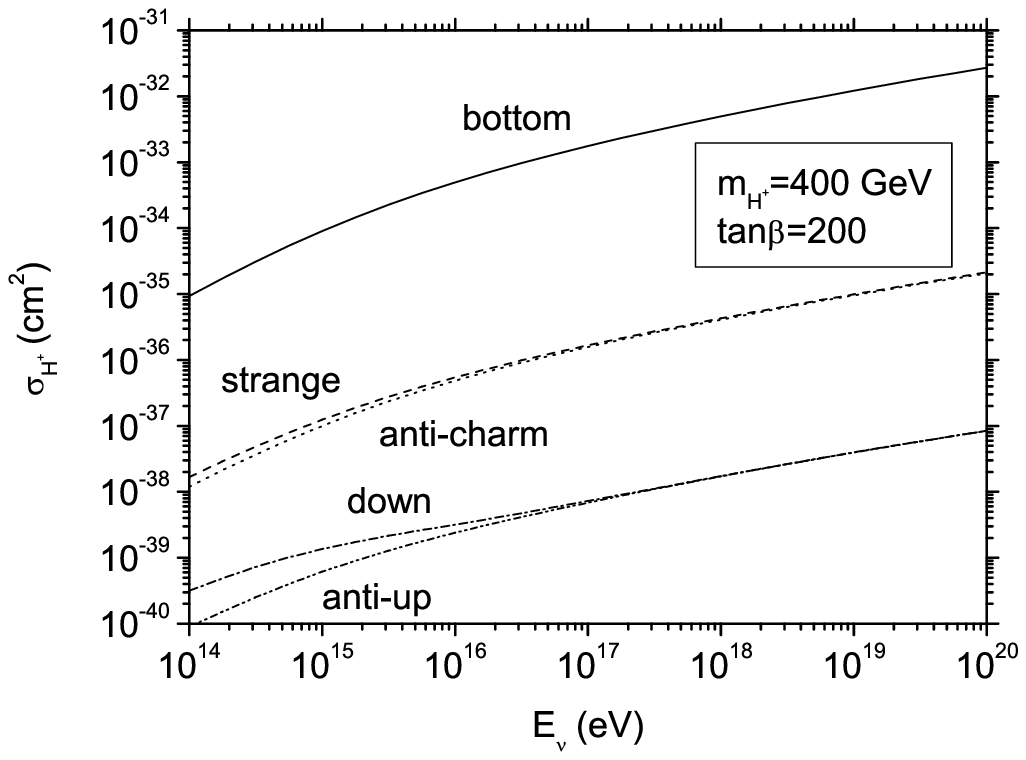}
\caption{Contribution to $\sigma^{tot}_{H^+}$ from the different
allowed initial quarks.}\label{fig:f3quarks}
\end{center}
\end{figure}

Finally, in Fig.~\ref{fig:f3quarks} we compare the contribution to
$\sigma^{tot}_{2hdm}(H^+)$ from the different allowed initial
quarks (see Fig.~\ref{fig:fig2hdm}(b)). We observe in this plot
that the contribution from the bottom quark dominates by far.

\section{Conclusions}

We have calculated the total cross section rates for the deep
inelastic process $\nu_{\tau} + {\cal N} \to \tau^- + X$, where
${\cal N} \equiv (n+p)/2$ is an isoscalar nucleon, in the frame of
the SM and the 2HDM(II). In the case of the 2HDM(II) we have taken
into account the contribution of the diagrams in which a charged
Higgs boson is exchanged $\sigma^{tot}_{H^+}$. We have shown that
the most important contribution to $\sigma^{tot}_{H^+}$ comes from
the $H^{\pm}$-exchange diagram with an initial b-quark (and hence
an outgoing t-quark). This fact implies that the contribution of
the $H^{\pm}$ exchange diagrams to the total cross section of the
$\nu_{\tau}\,{\cal N}$ scattering in the frame of the 2HDM is
independent whether the nucleon is a proton, a neutron or an
isoscalar nucleon, because these particles have the same content
of b-quark.

We showed that the contribution of the charged Higgs boson exchange
diagrams could imply an enhancement with respect to the SM cross
section rates for the charged current $\nu_{\tau} {\cal N}$ deep
inelastic scattering. We have obtained that for the case of an
ultrahigh energy tau-neutrino ($10^{14}$ eV $\leq E_{\nu} \leq$
$10^{20}$ eV) colliding on a target nucleon such enhancement can
reach up to $57\%$ and could help to discriminate between the SM and
the 2HDM predictions at the Pierre Auger Observatory.

\bigskip

\begin{center}
{\bf ACKNOWLEDGMENTS}
\end{center}
The author thanks {\it Sistema Nacional de Investigadores} and
{\it CONACyT} (M\'{e}xico) for financial support.

\end{document}